# Low-complexity implementation of convex optimization-based phase retrieval

Sercan Ö. Arık, and Joseph M. Kahn, *Fellow, IEEE*

*Abstract*— Phase retrieval has important applications in optical imaging, communications and sensing. Lifting the dimensionality of the problem allows phase retrieval to be approximated as a convex optimization problem in a higher-dimensional space. Convex optimization-based phase retrieval has been shown to yield high accuracy, yet its low-complexity implementation has not been explored. In this paper, we study three fundamental approaches for its low-complexity implementation: the projected gradient method, the Nesterov accelerated gradient method, and the alternating direction method of multipliers (ADMM). We derive the corresponding estimation algorithms and evaluate their complexities. We compare their performance in the application area of direct-detection mode-division multiplexing. We demonstrate that they yield small estimation penalties (less than 0.2 dB for transmitter processing and less than 0.6 dB for receiver equalization) while yielding low computational cost, as their implementation complexities all scale quadratically in the number of unknown parameters. Among the three methods, ADMM achieves convergence after the fewest iterations and the fewest computational operations.

## I. INTRODUCTION

PHASE retrieval corresponds to the extraction of phase information about signals solely from their measured intensities, possibly with the aid of various transformations.

Phase retrieval problems are commonly encountered in applications where phase measurements are impossible or impractical. One such field is optics, due to the fundamental difficulty of measuring phase at carrier frequencies of hundreds of THz [1]. The applications of phase retrieval in optics include imaging (crystallographic imaging [2-4], speckle imaging [5], coherent diffractive imaging [6,7], astronomical imaging [8,9]), sensing (wavefront sensing [11]), and communications (direct-detection mode-division multiplexing, MDM [10]). There are also important applications of phase retrieval outside of optics, such as multi-input multi-output (MIMO) wireless communications [12],
speech recognition [13] and compressed sensing [14-16].

The traditional setting of phase retrieval is in the recovery of signals from measured intensities in the space/spatial frequency or time/temporal frequency domains. This setting is encountered in many applications, because many systems (e.g., optical systems such as graded-index media, lenses, or simply free-space propagation) transform signals from the spatial domain into the spatial frequency domain, and many detection systems measure only intensities. The approaches typically used for phase retrieval in this traditional problem setting are of the form of sequential gradient descent. For example, the Gerchberg-Saxton algorithm is based on iteratively transforming a candidate solution back and forth between the two domains (e.g., space and spatial frequency), imposing constraints in one domain before transforming to the other [17,18]. Despite the good performance of such heuristic approaches, there is no algorithm for phase retrieval that guarantees achievement of a globally optimal solution at a low (non-deterministic polynomial-time hard) computational complexity. A fundamental challenge is the non-convexity of the intensity constraint after an orthogonal transformation [19].

Furthermore, in many applications, the problem setting for phase retrieval is more general than that addressed by traditional algorithms. An important generalization is encountered in recovery of unknown signals from the measured intensities of their inner products with known signals. (When the known signals are complex exponentials, we obtain the special case described in the previous paragraph.) The inner product operation occurs in linear systems when measurements are performed in the appropriate domain. For example, for a linear MIMO optical channel at a given frequency, the known signals may correspond to (baseband envelopes of) known training sequence vectors, and the signals to be recovered may correspond to rows of the channel transfer matrix [10]. In such settings, generalization of low-complexity heuristic approaches based on sequential gradient descent is challenging, as there might not be an orthogonal transformation between the known signals (e.g., Fourier transformation) with a low-complexity implementation (e.g., fast Fourier transform algorithm). The concept of sequential algorithms may be generalized in the form of alternating minimization, based on alternation between the missing phase information and the candidate solutions [20], but such methods do not reliably achieve good performance [2,10].

Addressing phase retrieval problems in a general setting,

This manuscript was submitted in September 2017. This work was supported in part by Huawei Technologies Co., Ltd.

S. Ö. Arık is with Baidu Silicon Valley Artificial Intelligence Lab, Sunnyvale, CA 94089 (e-mail: sercanarik@baidu.com).

J. M. Kahn is with the E. L. Ginzton Laboratory, Department of Electrical Engineering, Stanford University, Stanford, CA 94305 USA (e-mail: jmk@ee.stanford.edu).

Candes et al. recently proposed an approach of lifting the optimization problem to a high-dimensional space, where a convex optimization problem can efficiently approximate the original phase retrieval problem [21]. Convex optimization-based phase retrieval has been demonstrated to achieve high performance in imaging [21-23] and in communications [10]. Implementation complexity is critical in many applications, yet the low-complexity implementation of convex optimization-based phase retrieval has not yet been explored. For example, in short-reach optical communications [10], transceiver power consumption is an important concern, and can determine the practical feasibility of any proposed technology. Specialized low-complexity algorithms are required for such applications. Methods employed in general-purpose optimization software, such as interior-point methods, are typically not suitable for such applications.

In this paper, we study three fundamental methods for low-complexity implementation of convex optimization-based phase retrieval: the projected gradient method in Section III, the Nesterov accelerated gradient method in Section IV, and the alternating direction method of multipliers (ADMM) in Section V. For each method, we describe adaptive algorithms and present the associated computational complexities. In comparing the performance of these methods, we focus on an application area in optical communications: direct-detection MDM, whose architecture is described in detail in Section VI. We demonstrate that all these methods yield excellent performance at low implementation complexity in either transmitter- or receiver-based MIMO processing.

## II. CONVEX OPTIMIZATION-BASED PHASE RETRIEVAL

We focus on the problem of extracting unknown signal vectors from the magnitudes of their inner products with known signal vectors. We assume that $D$-dimensional known complex vectors $\mathbf{x}^{(n)}$ are used for estimation of $D$-dimensional unknown complex vectors $\mathbf{h}_l$. With additive noise $\mathbf{n}_l^{(n)}$, the available measurements are

$$\left\|\mathbf{h}_l^H \mathbf{x}^{(n)}\right\|^2 + \mathbf{n}_l^{(n)} = \mathbf{d}_l^{(n)}, \qquad (1)$$

where $1 \le n \le N$ and $1 \le l \le L$. We might have $L = D$, e.g., in direct-detection MDM for a $D \times D$ MIMO channel, or we might have $L \gg D$, e.g., in diffractive imaging when multiple illuminations are used.

In the absence of noise, the formulation (1) can be expressed equivalently as $L$ different rank-minimization problems. Defining $D \times D$ complex Hermitian matrix parameters $\mathbf{C}_l$, we have

$$\begin{aligned}\min \quad & \text{rank}(\mathbf{C}_l) \\ \text{s.t.} \quad & \mathbf{C}_l \ge 0 \text{ and } \mathbf{x}^{(n)H}\mathbf{C}_l\mathbf{x}^{(n)} = \mathbf{d}_l^{(n)}, \text{ for } 1 \le n \le N\end{aligned}. \quad (2)$$

Optimal solutions to (2) are obtained for the matrix parameters $\hat{\mathbf{C}}_l = \mathbf{h}_l^H \mathbf{h}_l$. To consider the effect of noise, the equality constraints in (2) can be expressed as an additive penalty term in a maximum-likelihood formulation of signal retrieval with additive noise. In this paper, we focus on spatially white zero-mean real-valued Gaussian noise, for which the rank-minimization problem can be expressed as

$$\begin{aligned}\min \quad & \text{rank}(\mathbf{C}_l) + \lambda \sum_{n=1}^{N} \left\|\mathbf{x}^{(n)H}\mathbf{C}_l\mathbf{x}^{(n)} - \mathbf{d}_l^{(n)}\right\|^2, \\ \text{s.t.} \quad & \mathbf{C}_l \ge 0\end{aligned} \quad (3)$$

where $\lambda$ is a tradeoff constant and $\ge$ denotes that the matrix is positive semidefinite. Generalization of (3) to other noise types is possible by considering the maximum-likelihood formulation for the corresponding noise statistics, which might or might not be convex in $\mathbf{x}^{(n)}$.

As rank-minimization problems are non-deterministic polynomial-time hard, it is desired to apply a convex relaxation that can approximate the optimization problem (3) efficiently, while yielding a global optimal solution. An efficient convex relaxation is based on substituting the trace function [21] in place of the rank function, which yields

$$\begin{aligned}\min \quad & g(\mathbf{C}_l) = \text{tr}(\mathbf{C}_l) + \lambda \sum_{n=1}^{N} \left\|\mathbf{x}^{(n)H}\mathbf{C}_l\mathbf{x}^{(n)} - \mathbf{d}_l^{(n)}\right\|^2. \\ \text{s.t.} \quad & \mathbf{C}_l \ge 0\end{aligned} \quad (4)$$

After computing the optimal matrix parameter $\hat{\mathbf{C}}_l$, its first principal component $\sqrt{\hat{\chi}_{1,1}}\hat{\mathbf{u}}_{1,1}$, where $\hat{\mathbf{C}}_l = \sum_{k=1}^{D} \hat{\chi}_{i,k} \hat{\mathbf{u}}_{i,k} \hat{\mathbf{u}}_{i,k}^H$, yields the estimate for $\mathbf{h}_l$. The overall dimensionality of the problem is squared in exchange for the benefit of convex approximation.

We will next focus on three techniques that aim to approximately solve the increased-dimensionality problem efficiently with low computational complexity. In estimating computational complexity, we assume $\Re(\ )$, $\Im(\ )$ and $(\ )^*$ operations have negligible cost. We quantify the computational complexity of other operations, which are generally performed on complex numbers, in terms of the number of floating point operations (flops) required. We assume optimal ordering of algebraic operations to minimize complexity. For example, multiplication of a matrix, vector and scalar can be done with lower complexity by first multiplying the matrix and the vector and then the scalar, or by first multiplying the vector and the scalar and then the matrix, as opposed to first multiplying the scalar and the matrix and then the vector. We also assume that any computed result can be reused at multiple steps without being recomputed, by utilizing a working memory that can store intermediate results. The structure of the objective function in (4) forms the basis for the reduced-complexity algorithms.

## III. PROJECTED GRADIENT METHOD

Projected gradient methods represent an efficient first-order approach for constrained convex optimization problems. The principle is based on iteratively searching for optimal values of the variables along a direction determined by gradient values, while imposing the constraints at each step.

Starting with an initial guess $\mathbf{C}_l[0]$, the projected gradient method applies the following update equation recursively at iteration step $k+1$:

$$\mathbf{C}_l[k+1] \leftarrow P(\mathbf{C}_l[k] - t[k]\nabla g(\mathbf{C}_l[k])). \quad (5)$$

In (5), $\nabla g(\ )$ denotes the gradient operator corresponding to the objective function and $P(\ )$ denotes the projection operator that is required to satisfy the constraints.

We will next derive the various operators and analyze their computational complexities. The step size can be updated by a search algorithm. A common approach is a back-tracking line search algorithm:

**while** ( $g(\mathbf{C}_l[k]) < g(\mathbf{C}_l[k] - t[k] \nabla g(\mathbf{C}_l[k]))$ ):
$$\textbf{do } t[k] = t[k] \cdot \tau \quad (6)$$

where $0 < \tau < 1$. The computational complexity of back-tracking line search is dominated by the complexity of the objective function computation, which is $O(D^2 LN)$ flops per iteration step.

Computation of $\nabla g(\mathbf{C}_l)$, the gradient with respect to $\mathbf{C}_l$, requires computing the following derivatives for $1 \leq i \leq D$ and $i+1 \leq k \leq D$:

$$\frac{\partial g(\mathbf{C})}{\partial \Re(\mathbf{C}_{lii})} = 1 + 2\lambda \left( \sum_{n=1}^{N} \left| \mathbf{x}_i^{(n)} \right|^2 \Re\left(\mathbf{x}^{(n)H} \mathbf{C}_l \mathbf{x}^{(n)}\right) \right) \\ - 2\lambda \left( \sum_{n=1}^{N} \Re\left(\mathbf{d}_l^{(n)} \left| \mathbf{x}_i^{(n)} \right|^2 \right) \right) \quad (7)$$

$$\frac{\partial g(\mathbf{C})}{\partial \Re(\mathbf{C}_{lik})} = 2\lambda \Re\left( \sum_{n=1}^{N} \mathbf{x}_i^{(n)} \mathbf{x}_k^{(n)*} \left(\mathbf{x}^{(n)H} \mathbf{C}_l \mathbf{x}^{(n)}\right) \right) \\ - 2\lambda \Re\left( \sum_{n=1}^{N} \mathbf{d}_l^{(n)} \mathbf{x}_i^{(n)} \mathbf{x}_k^{(n)*} \right) \quad (8)$$

$$\frac{\partial g(\mathbf{C})}{\partial \Im(\mathbf{C}_{lik})} = 2\lambda \Im\left( \sum_{n=1}^{N} \mathbf{x}_i^{(n)} \mathbf{x}_k^{(n)*} \left(\mathbf{x}^{(n)H} \mathbf{C}_l \mathbf{x}^{(n)}\right) \right) \\ - 2\lambda \Im\left( \sum_{n=1}^{N} \mathbf{d}_l^{(n)} \mathbf{x}_i^{(n)} \mathbf{x}_k^{(n)*} \right) \quad (9)$$

Computation of the $\left|\mathbf{x}_i^{(n)}\right|^2$ and $\mathbf{x}_i^{(n)*} \mathbf{x}_k^{(n)}$ terms require $O(ND^2)$ flops in total. For all expressions in (7)-(9), the $\mathbf{x}^{(n)H} \mathbf{C}_l \mathbf{x}^{(n)}$ terms are computed once with a complexity of $O(D^2 L)$ flops. Given $\left|\mathbf{x}_i^{(n)}\right|^2$, $\mathbf{x}_i^{(n)*} \mathbf{x}_k^{(n)}$, and $\mathbf{x}^{(n)H} \mathbf{C}_l \mathbf{x}^{(n)}$ terms, computation of $\sum_{n=1}^{N} \mathbf{x}_i^{(n)} \mathbf{x}_k^{(n)*} \left(\mathbf{x}^{(n)H} \mathbf{C}_l \mathbf{x}^{(n)}\right)$ or $\sum_{n=1}^{N} \left|\mathbf{x}_i^{(n)}\right|^2 \Re\left(\mathbf{x}^{(n)H} \mathbf{C}_l \mathbf{x}^{(n)}\right)$ require $2N-1$ flops for each $i$, $k$ and $l$, yielding a total complexity of $O(D^2 LN)$ flops. To construct all the derivatives, performing all the additions and subtractions in (7)-(9) requires $O(D^2 L)$ flops. The total complexity of computing all the derivatives is $O(D^2 LN)$ flops.

Imposing the constraint is based on the projection operator onto the positive-semidefinite cone:

$$P\left( \sum_{i=1}^{D} \lambda_i \mathbf{u_i} \mathbf{u_i}^H \right) = \sum_{i=1}^{D} \max(0, \lambda_i) \mathbf{u_i} \mathbf{u_i}^H . \quad (10)$$

Computation of the projection operator requires eigenvalue decomposition. Hardware-efficient eigenvalue decomposition techniques depend on the size of the matrix. For $D = 3$, techniques based on the analytical expressions of the eigenvalues and eigenvectors are fastest [1]. For $D > 25$, divide-and-conquer is the fastest algorithm [2]. For $3 < D < 25$, (the range of interest for practical applications, e.g., in MDM systems), tridiagonal QR iteration is the fastest algorithm, with a complexity of $O(D^3)$ flops [24]. After replacing any negative eigenvalues by 0, the reconstruction of the projection requires $O(D^3)$ flops. Overall, applying projection $L$ times requires in total $O(D^3 L)$ flops per iteration,

For the projected gradient method, the overall computational cost is dominated by the computation of gradient and projection operations, and is $\max\left(O(D^2 LN), O(D^3 L)\right)$ flops per iteration. The total computational complexity is also proportional to the number of iterations. Hence, minimization of the total number of iterations is also crucial.

## IV. NESTEROV ACCELERATED GRADIENT METHOD

Optimal first-order methods are derived from classical projected gradient methods, with the goal of reducing the number of iterations required to achieve convergence.

One modification of the projected gradient method, pioneered by Nesterov [25], is the accelerated gradient method [26], which has update equations:

$$\mathbf{C}_l[k+1] \leftarrow P(\mathbf{K}_l[k] - t[k+1] \nabla g(\mathbf{K}_l[k])), \quad (11)$$

$$\gamma[k+1] \leftarrow \frac{1}{\left(\frac{1}{2} + \sqrt{\frac{1}{4} + \frac{1}{\gamma[k]^2}}\right)}, \quad (12)$$

$$\mathbf{K}_l[k+1] \leftarrow \mathbf{C}_l[k+1] + \frac{\gamma[k+1](1-\gamma[k])}{\gamma[k]} \left(\mathbf{C}_l[k+1] - \mathbf{C}_l[k]\right). \quad (13)$$

For determining the step size in (11), as above, the back-tracking line search algorithm (6) can be used. The fundamental distinction of the Nesterov accelerated gradient method is the adjustment of the step size using the $\gamma[k]$ parameter. The evolution of $\gamma[k]$ as a function of the iteration number is shown in Fig. 1.

The Nesterov accelerated gradient method has a slightly higher complexity than the projected gradient method. The computational cost for the update step (12) is $O(1)$ flops and for (13) is $O(D^2 L)$ flops. The major computational cost is step (11), which requires $\max\left(O(D^2 LN), O(D^3 L)\right)$ flops per iteration, similar to the projected gradient algorithm. Again, the total computational complexity is proportional to the number of iterations. The modification for acceleration reduces the number of steps required for convergence, despite a slight increase in the computational cost per iteration step.

## V. ALTERNATING DIRECTION METHOD OF MULTIPLIERS

A general approach to address constrained convex optimization problems with complicated objective functions is

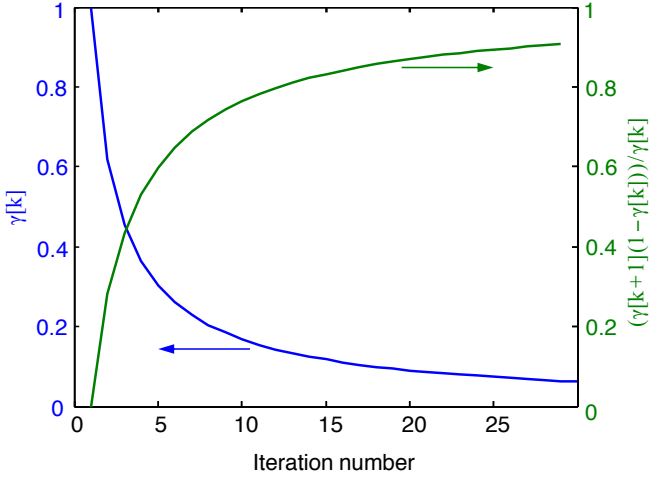

**Fig. 1** $\gamma[k]$ vs. iteration number for the Nesterov accelerated projected gradient method.

to decompose them into smaller problems, each of which is then easier to handle. The ADMM technique combines the benefits of the dual decomposition and augmented Lagrangian methods for constrained optimization,

We start by observing that the optimization problem can be equivalently expressed in equality constraint form:

$$\begin{aligned} \min \quad & g(\mathbf{C}_l) + I_{\geq 0}(\mathbf{D}_l) \\ \text{s.t.} \quad & \mathbf{C}_l = \mathbf{D}_l \end{aligned} \quad (14)$$

where $I_{\geq 0}(\mathbf{D}_l)$ is the indicator function such that

$$I_{\geq 0}(\mathbf{D}_l) = \begin{cases} 0, & \mathbf{D}_l \geq 0 \\ \infty, & \text{otherwise} \end{cases}. \quad (15)$$

The corresponding augmented Lagrangian is in the form

$$L(\mathbf{C}_l, \mathbf{D}_l, \mathbf{E}_l) = g(\mathbf{C}_l) + I_{\geq 0}(\mathbf{D}_l) + \langle \mathbf{E}_l, \mathbf{D}_l - \mathbf{C}_l \rangle + \rho \|\mathbf{D}_l - \mathbf{C}_l\|_F^2. \quad (16)$$

ADMM is based on replacing minimization of (16) by alternating minimization, where each step optimizes the matrix parameters $\mathbf{C}_l$, $\mathbf{D}_l$ and $\mathbf{E}_l$ separately:

$$\mathbf{C}_l[k+1] \leftarrow \arg\min_{\mathbf{C}_l} L(\mathbf{C}_l, \mathbf{D}_l[k], \mathbf{E}_l[k]), \quad (17)$$

$$\mathbf{D}_l[k+1] \leftarrow \arg\min_{\mathbf{D}_l} L(\mathbf{C}_l[k+1], \mathbf{D}_l, \mathbf{E}_l[k]), \quad (18)$$

$$\mathbf{E}_l[k+1] \leftarrow \mathbf{E}_l[k] + \rho(\mathbf{D}_l[k+1] - \mathbf{C}_l[k+1]). \quad (19)$$

We now analyze the steps (17)-(19).

For step (17), we find the analytical solution by setting the first partial derivatives with respect to $\mathbf{C}_{lik}$ to zero (since the second derivatives are always positive), which yield

$$\rho \mathbf{C}_{lik}[k+1] + \sum_{a,b=1}^{D} \left( \mathbf{C}_{lab}[k+1] \mathbf{T}_{ikab}^{(N)} \right) = \\ \mathbf{U}_{lik}^{(N)} - \left( \frac{\mathbf{E}_{lik}[k]}{2} + \rho \mathbf{D}_{lik}[k] \right) - \frac{\delta_{i,k}}{2}, \quad (20)$$

where $\delta_{i,k}$ is a Kronecker delta function, and

$$\mathbf{U}_{lik}^{(N)} = \lambda \sum_{n=1}^{N} \mathbf{d}_l^{(n)} \mathbf{x}_i^{(n)} \mathbf{x}_k^{(n)*}, \quad (21)$$

$$\mathbf{T}_{ikab}^{(N)} = \lambda \sum_{n=1}^{N} \mathbf{x}_i^{(n)} \mathbf{x}_k^{(n)*} \mathbf{x}_b^{(n)} \mathbf{x}_a^{(n)*}. \quad (22)$$

The minimization problem (17) reduces to a set of $D^2$ linear equations. Consider the matrix representation of the set of linear equations (20)

$$\left( \widetilde{\mathbf{X}}^{(N)} + \rho \mathbf{I} \right) \begin{bmatrix} \mathbf{C}_{l11}[k+1] \\ \mathbf{C}_{l12}[k+1] \\ \vdots \\ \mathbf{C}_{lD(D-1)}[k+1] \\ \mathbf{C}_{lDD}[k+1] \end{bmatrix} = \mathbf{b}_l^{(N)}[k], \quad (23)$$

where $\mathbf{b}_l^{(N)}[k]$ is a vector of size $D^2 \times 1$ such that $\mathbf{b}_{l((i-1)D+k)}^{(n)}[k] = \mathbf{U}_{lik}^{(N)} - (\mathbf{E}_{lik}[k]/2 + \rho \mathbf{D}_{lik}[k]) - \delta_{i,k}/2$, $\widetilde{\mathbf{X}}^{(N)} = \lambda \sum_{n=1}^{N} \mathbf{X}^{(n)} \mathbf{X}^{(n)T}$ is a matrix of size $D^2 \times D^2$, and $\mathbf{X}_{(i-1)D+k}^{(n)} = \mathbf{x}_i^{(n)*} \mathbf{x}_k^{(n)}$ is a vector of size $D^2 \times 1$. For the case of invertible $(\widetilde{\mathbf{X}}^{(N)} + \rho \mathbf{I})$, the solution of (23) is given by

$$\begin{bmatrix} \mathbf{C}_{l11}[k+1] \\ \mathbf{C}_{l12}[k+1] \\ \vdots \\ \mathbf{C}_{lD(D-1)}[k+1] \\ \mathbf{C}_{lDD}[k+1] \end{bmatrix} = \left( \widetilde{\mathbf{X}}^{(N)} + \rho \mathbf{I} \right)^{-1} \mathbf{b}_l^{(N)}[k] = \mathbf{F}^{(N)} \mathbf{b}_l^{(N)}[k]. \quad (24)$$

Inverting an ordinary matrix of size $D^2 \times D^2$ requires $O(D^6)$ flops using naïve algorithms such as Gauss-Jordan elimination, and requires $O(D^{4.746})$ flops with optimized Coppersmith-Winograd-like construction methods [27]. Iterative computation of the matrix inverse term $\mathbf{F}^{(N)}$ is also possible using the Sherman–Morrison–Woodbury lemma, exploiting its structure:

$$\begin{aligned} \mathbf{F}^{(N)} &= \left( \widetilde{\mathbf{X}}^{(N)} + \rho \mathbf{I} \right)^{-1} \\ &= \left( \lambda \mathbf{X}^{(N)} \mathbf{X}^{(N)T} + \lambda \sum_{n=1}^{N-1} \mathbf{X}^{(n)} \mathbf{X}^{(n)T} + \rho \mathbf{I} \right)^{-1} \quad (25) \\ &= \mathbf{F}^{(N-1)} - \frac{\mathbf{F}^{(N-1)} \mathbf{X}^{(N)} \mathbf{X}^{(N)T} \mathbf{F}^{(N-1)}}{1/\lambda + \mathbf{X}^{(N)T} \mathbf{F}^{(N-1)} \mathbf{X}^{(N)}} \end{aligned}$$

Note that we have the initial conditions $\widetilde{\mathbf{X}}^{(0)} = \mathbf{0}$ and $\mathbf{F}^{(0)} = \mathbf{I}/\rho$. Computation of $\mathbf{X}^{(N)}$ requires $O(D^2)$ flops. Given $\mathbf{X}^{(N)}$ and $\mathbf{F}^{(N-1)}$, the update of $\mathbf{F}^{(N)}$ requires $O(D^4)$ flops. As the update is repeated for all $n$, the total complexity of computing $\mathbf{F}^{(N)}$ becomes $O(D^4 N)$ flops. Computation of all $\mathbf{U}_{lik}^{(N)}$ terms requires an additional $O(D^2 NL)$ flops. Given $\mathbf{U}_{lik}^{(N)}$, computation of all $\mathbf{b}_l^{(N)}[k]$ vectors requires $O(D^2 L)$ flops. Overall, computation of $\mathbf{C}_l[k+1]$ requires $\max(O(D^2 LN), O(D^4 N))$ flops at each iteration.

For step (18), an analytical solution can be obtained by equating the gradient to zero. We have the gradient

$$\nabla_{\mathbf{D}_l[k]}\left(\left\|\frac{1}{2\rho}\mathbf{E}_l[k] + (\mathbf{D}_l[k] - \mathbf{C}_l[k+1])\right\|_F^2\right) = \\ 2\left(\frac{1}{2\rho}\mathbf{E}_l[k] + (\mathbf{D}_l[k] - \mathbf{C}_l[k+1])\right) \quad (26)$$

If $\mathbf{C}_l[k+1] - \frac{1}{2\rho}\mathbf{E}_l[k] \geq 0$, then the optimal solution is $\mathbf{D}_l[k+1] = \mathbf{C}_l[k+1] - \frac{1}{2\rho}\mathbf{E}_l[k]$. For the general case, the optimal $\mathbf{D}_l[k+1]$ is the projection $P\left(\mathbf{C}_l[k+1] - \frac{1}{2\rho}\mathbf{E}_l[k]\right)$. The overall update equation becomes

$$\mathbf{D}_l[k+1] \leftarrow P\left(\mathbf{C}_l[k+1] - \frac{1}{2\rho}\mathbf{E}_l[k]\right). \quad (27)$$

Computation of the $\mathbf{C}_l[k+1] - \frac{1}{2\rho}\mathbf{E}_l[k]$ terms requires $O(D^2 L)$ flops in total. The projection operator is applied $L$ times and dominates the computational complexity, requiring $O(D^3 L)$ flops per iteration.

Lastly, step (19) requires $O(D^2 L)$ flops at each step. The overall computational complexity per step is $\max(O(D^2 LN), O(D^4 N), O(D^3 L))$ flops per iteration, and is again proportional to the total number of iterations.

The computational complexities of the three methods are summarized in Table I.

TABLE I
COMPUTATIONAL COMPLEXITIES OF PHASE RETRIEVAL METHODS

|  | Computational complexity per iteration |
|---|---|
| Projected gradient | $\max(O(D^2 LN), O(D^3 L))$ |
| Nesterov | $\max(O(D^2 LN), O(D^3 L))$ |
| ADMM | $\max(O(D^2 LN), O(D^4 N), O(D^3 L))$ |

## VI. APPLICATION EXAMPLE: DIRECT-DETECTION MODE-DIVISION MULTIPLEXING

In this section, we study application of the phase retrieval techniques described above in direct-detection MDM [10].

MDM aims to increase the data throughput per fiber by employing a fiber supporting $D$ propagating modes. At the transmitter, $D$ independent data signals are multiplexed onto these modes, ideally providing a $D$-fold increase in throughput. During transmission, the data signals are intermixed by random coupling between modes, which corresponds to a $D \times D$ channel transfer matrix that has off-diagonal entries. We assume that group velocity dispersion and modal dispersion are negligible, such that memoryless signal processing and channel estimation are sufficient.

Using direct detection simplifies receiver implementation, making the technique potentially suitable for short-reach communications. However, only the intensity (squared modulus) of each modal field amplitude is available at the receiver. This has important consequences for MIMO signal processing, channel transfer matrix estimation and information decoding. First, compensating for modal crosstalk requires complex-valued $D \times D$ MIMO signal processing. As shown in Fig. 2, direct-detection MDM can be implemented using a digital MIMO precoder or optical MIMO precoder at the transmitter, or using an optical MIMO equalizer at the receiver (additional details are given in [10]). Second, adjustment of the MIMO precoder or equalizer requires estimation of the $D \times D$ channel transfer matrix using phase retrieval. Third, information must be demodulated from the optical carrier using noncoherent or differentially coherent detection.

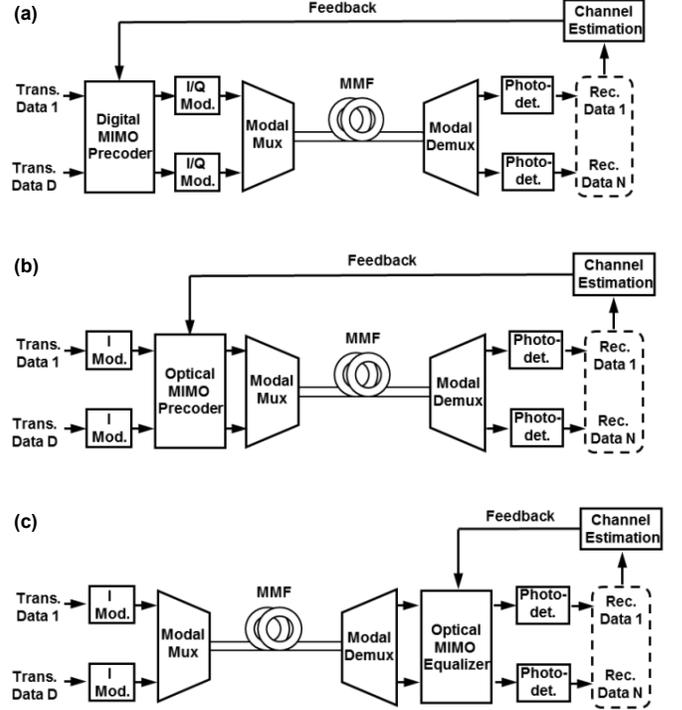

**Fig. 2** Direct detection MDM systems using: (a) digital MIMO precoding at the transmitter, (b) optical MIMO precoding at the transmitter, and (c) optical MIMO equalization at the receiver. Mux: Multiplexer, mod.: modulator, det.: detector, I: in-phase, Q: quadrature.

Here, we are concerned with the second consequence, namely, how to perform channel estimation using phase retrieval. Estimation is facilitated using a known training sequence, which is composed of $N$ $D$-dimensional signals. For a given $D$, the required size of the training signals can be found empirically. If the number of training vectors $N$ is not large enough, high estimation error is obtained due to insufficient measurement diversity. On the other hand, increasing $N$ beyond a certain threshold does not improve estimation accuracy significantly, and it is desired to set the value of $N$ around this threshold. Note that for optical precoding or equalization methods, the training vectors can be transmitted at a low symbol rate, and high-speed digital signal processing can be avoided. For all three methods (digital or optical MIMO precoding or optical equalization), channel estimation corresponds to the phase retrieval problems described in Section 2.

For a MIMO precoder, the channel transfer matrix needs to be estimated up to a constant diagonal unitary matrix [10]. We have $L = D$, since for each transmitted sequence, $D$ different inner products are generated by the $D$ rows of the channel transfer matrix, i.e., $\mathbf{h}_l$ in (1) corresponds to the $l^{th}$ row of the channel transfer matrix.

For a MIMO equalizer, the channel transfer matrix needs to be estimated up to a complex exponential, hence an additional optimization procedure is used to estimate the unknown phases of each row of the channel transfer matrix, yielding $L = D+1$ [10].

Considering $L = D$ or $L = D+1$ with Table I, for a MIMO precoder or a MIMO equalizer, we estimate the computational complexity per iteration to be $\max(O(D^3 N), O(D^4))$ for the projected gradient and Nesterov accelerated gradient methods, and $O(D^4 N)$ for the ADMM. In other words, they are of the order of the dimensionality of the high-dimensional space, which is the square of the number of unknowns.

In short-reach systems, the end-to-end link may be optimized to have negligible modal crosstalk between mode groups [28]. In this case, MIMO processing can be implemented within each mode group separately. Fig. 3 illustrates processing in mode group subspaces in a system using receiver MIMO equalization implemented by an array of Mach-Zehnder interferometers [29]. Processing in mode group subspaces can reduce the total hardware complexity of MIMO processing, as the required number of Mach-Zehnder interferometers scales with the square of the number of modes processed. In addition, mode group subspace processing can significantly reduce the computational complexity of channel estimation, as the estimation algorithms can be implemented separately for each mode group (e.g., of sizes $D = 2$ and $D = 4$ for the system shown in Fig. 3).

## VII. NUMERICAL RESULTS

In the simulations presented here, we consider a direct-detection MDM system employing on-off keying. We assume the receiver is thermal noise-limited, so the received modal intensities are corrupted by additive, spatially white real Gaussian noise. Realizations of the channel transfer matrix are chosen from an ensemble of fully random complex-valued $D \times D$ unitary matrices to model worst-case modal crosstalk. We assume $D = 6$ modes and, accordingly, we choose $N = 300$ training signals to obtain low estimation error. We assume that these $N$ training signals are generated randomly, with each vector element an independent, identically distributed circularly symmetric Gaussian random variable. The convex optimization formulation tradeoff constant is chosen as $\lambda = 10$, as in [10], as this choice yielded the lowest estimation error in the original convex optimization problem formulation (4). For projected gradient and Nesterov accelerated gradient methods, the backtracking line search parameter is chosen as $\tau = 0.3$, which is at an optimal point when the tradeoff between the number of backtracking line search iterations and the direction of search accuracy is considered (when the value of $\tau$ is close to 1, convergence of the backtracking line search would be slow and when the value of $\tau$ value is too small, the estimation error would be high when moving in the direction of the step size).

The system electrical signal-to-noise ratio (SNR) is defined as

$$SNR = \left[\left\langle \sum_{i=1}^{D} |\mathbf{y}_i^{(n)}|^2 \right\rangle \Big/ (D \cdot \sigma) \right]^2, \quad (28)$$

where $\langle \ \rangle$ denotes an average over training signals. Since (28) is proportional to the square of the optical power per mode, a 1-dB change of the optical power per mode corresponds to a 2-dB change of $10\log_{10}(SNR)$. A range of SNR values is considered that corresponds to practical short-reach optical links. Link performance is quantified by the average bit-error ratio (BER). Note that short-reach communication systems typically use forward error-correction codes with low complexity, necessitating BER targets around $10^{-3}$ to $10^{-4}$.

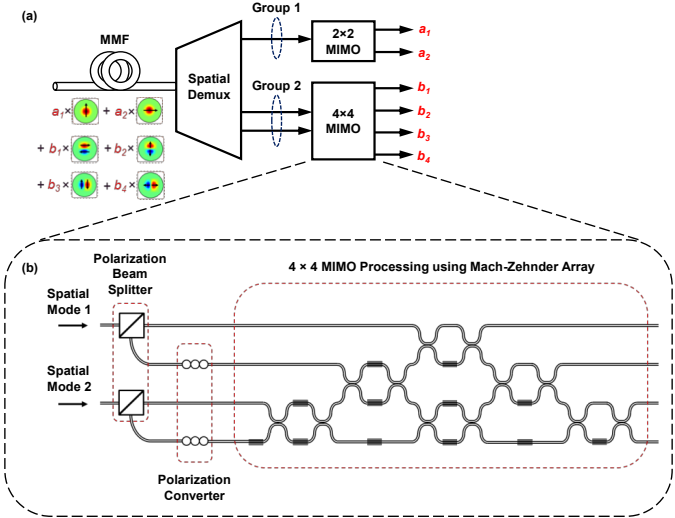

Fig. 3 (a) Demultiplexing using receiver MIMO processing in mode group subspaces, (b) implementation of optical MIMO processing using Mach-Zehnder array.

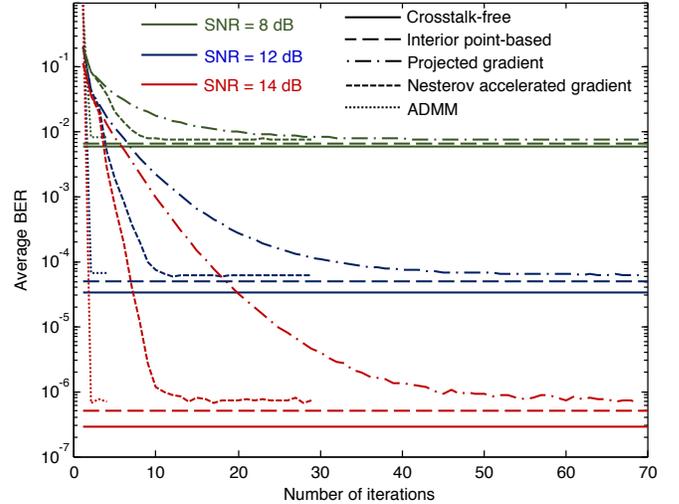

**Fig. 4** Average BER vs. number of iterations for a direct-detection MDM system using on-off keying, transmitter MIMO precoding, $D = 6$ modes and $N = 300$ training signals. Horizontal lines show the performance achieved in the two reference cases: for the ideal crosstalk-free channel and when an interior-point technique [28] is used to implement convex estimation-based phase retrieval. ADMM: alternating direction method of multipliers.

We compare the performance of the three low-complexity convex optimization-based phase retrieval techniques to the performance achieved in two reference cases. The first reference case is the ideal crosstalk-free channel, where additive noise is the only impairment. The second reference case includes modal crosstalk and uses CVX [30], a well-known interior-point-based package, for convex estimation-based phase retrieval to determine the MIMO precoder or equalizer.

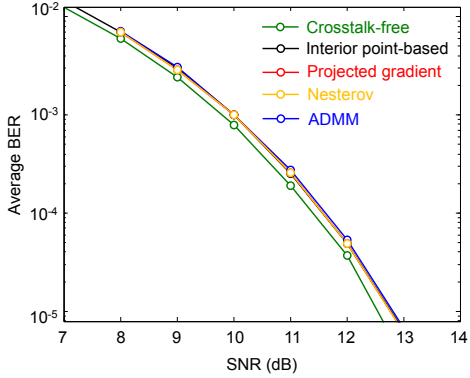

**Fig. 5** Average BER vs. SNR for a direct-detection MDM system using on-off keying, transmitter MIMO precoding, $D = 6$ modes and $N = 300$ training signals. ADMM: alternating direction method of multipliers.

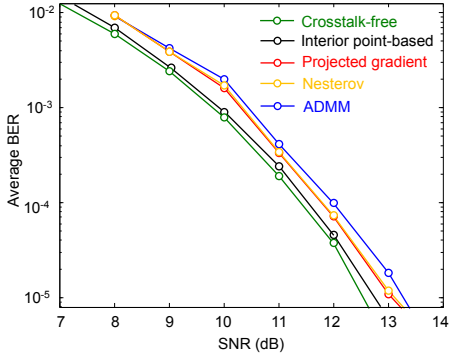

**Fig. 6** Average BER vs. SNR for a direct-detection MDM system using on-off keying, receiver MIMO equalization, $D = 6$ modes and $N = 300$ training signals. ADMM: alternating direction method of multipliers.

Fig. 4 shows the average BER vs. number of iterations using transmitter MIMO precoding. The convergence speed is rather insensitive to the SNR value. The projected gradient method converges after ~50 iterations, the Nesterov accelerated gradient method converges after ~10 iterations, and the ADMM converges after just a few iterations. In the remainder of the simulations, we fix the total number of iterations to 70 for the projected gradient method, 30 for the Nesterov accelerated gradient method and 5 for the ADMM, to ensure convergence beyond the respective knees of the curves shown in Fig. 4.

Table II provides estimates of the number of operations per second required for channel estimation based on the assumptions described in Sections III-V. We note that the mapping of an algorithm to hardware can be optimized given tradeoffs between the costs of computation, memory, and communication on a chip. Hence, the numerical values in Table II should be considered as rough order-of-magnitude estimates for straightforward hardware mapping on a single-core processor without parallelization of the computation.

The operation counts in Table II are computed assuming channel estimation at intervals of 1, 10 and 100 ms. Based on studies of channel dynamics [31], we estimate that short-reach multimode links with D~6 modes will change on time scales of ms to tens of ms [31]. Even if a direct-detection MDM system must update channel estimates at 1 ms intervals, the operation count for ADMM is only ~5 Gflop/s. This is about two orders of magnitude smaller than the operation count for a long-haul coherent MDM receiver [28].

TABLE II
OPERATIONS PER SECOND FOR CONVEX OPTIMIZATION-BASED PHASE RETRIEVAL IN A DIRECT-DETECTION MDM SYSTEM USING TRANSMITTER MIMO PRECODING, ASSUMING $L = D = 6$, $N = 300$.

|  | Operations per second required for channel estimation for update intervals indicated | | |
|---|---|---|---|
| Update interval | 1 ms | 10 ms | 100 ms |
| Projected gradient | ~50 Glops/s | ~5 Gflops/s | ~500 Mflops/s |
| Nesterov | ~20 Gflops/s | ~2 Gflops/s | ~200 Mflops/s |
| ADMM | ~5 Gflops/s | ~0.5 Gflops/s | ~50 Mflops/s |

Fig. 5 shows the average BER vs. SNR using transmitter MIMO precoding. All three techniques exhibit performance very close to CVX, incurring SNR penalties (with respect to crosstalk-free case) less than 0.3 dB for target average BERs of $10^{-3}$ to $10^{-4}$.

Fig. 6 shows the average BER vs. SNR using receiver MIMO equalization. Receiver MIMO equalization necessitates estimation procedures to determine the transmitter MIMO precoder, as well as additional estimation to disambiguate common phase factors [10]. Consequently, there are more opportunities for estimation errors than in transmitter precoding. The projected gradient and Nesterov accelerated gradient methods yield similar performance, incurring SNR penalties (with respect to crosstalk-free case) less than 0.6 dB for target average BERs of $10^{-3}$ to $10^{-4}$. For ADMM, the SNR penalty is slightly higher, but still less than 0.8 dB for the same target average BER range.

## VIII. CONCLUSIONS

We have studied three low-complexity approaches for implementing convex optimization-based phase retrieval: the projected gradient method, the Nesterov accelerated gradient method and ADMM. The projected gradient method is a first-order technique, which aims to impose the constraint at each iteration step. The Nesterov accelerated gradient method achieves an acceleration of the projected gradient method by using an adaptive step size parameter. The ADMM is based on redefining the convex optimization objective for uncoupled solutions of the separable objective functions using alternating minimization. We have demonstrated application of the three techniques to direct-detection MDM. All three methods accurately approximate the solution to the original phase retrieval problem, incurring small estimation penalties. The computational complexities of all three methods scale with the square of the number of unknowns per iteration. Among the three, ADMM yields convergence after the smallest number of iterations.

ACKNOWLEDGEMENTS

Discussions with Prof. Emmanuel J. Candes are gratefully acknowledged.